\documentclass[preprint,showpacs,aps,nofootinbib,superscriptaddress]{revtex4-1}
\usepackage{amssymb}
\usepackage{amsmath,amssymb,graphicx}
\usepackage{graphicx}
\usepackage{dcolumn}
\usepackage{bm}

\def\beq{\begin{equation}}
\def\eeq{\end{equation}}
\def\bea{\begin{eqnarray}}
\def\eea{\end{eqnarray}}

\usepackage{pstricks}
\usepackage{graphicx}
\usepackage{amsbsy}
\usepackage{bm}
\usepackage{color}
\usepackage{fancyhdr}
\usepackage{epsfig}
\usepackage[colorlinks=true,linkcolor=blue,citecolor=red,urlcolor=green,filecolor=blue]{hyperref}
\usepackage{slashed}
\usepackage{multirow}

\begin{document}

\bigskip

\vspace{2cm}
\title{Sensitivity to Majorana neutrinos in $\Delta L= 2$ decays of $B_c$ meson at LHCb}

\vskip 6ex
\author{Diego Milan\'{e}s}
\email{diego.milanes@cern.ch}
\affiliation{Departamento de F\'{i}sica, Universidad Nacional de Colombia, C\'{o}digo Postal 11001, Bogot\'{a} D.C., Colombia}
\author{N\'{e}stor Quintero}
\email{nhquinterop@ut.edu.co}
\affiliation{Departamento de F\'{i}sica, Facultad de Ciencias, Universidad del Tolima, C\'{o}digo Postal 730006299, Ibagu\'{e}, Colombia}
\author{Carlos E. Vera}
\email{cvera@ut.edu.co}
\affiliation{Departamento de F\'{i}sica, Facultad de Ciencias, Universidad del Tolima, C\'{o}digo Postal 730006299, Ibagu\'{e}, Colombia}

\bigskip\bigskip
\begin{abstract}
The possible existence of Majorana neutrinos can be tested through the study of processes  where the total lepton number $L$ is violated by two units ($\Delta L=2$). In this work, the production of an on-shell Majorana neutrino with a mass around $\sim$ 0.2 GeV to a few GeV is studied in $\Delta L= 2$ decays of the $B_c$ meson. We focus on the same-sign di-muon channels: three-body $B_c^- \to \pi^+\mu^-\mu^-$ and four-body $B_c^- \to J/\psi \pi^+\mu^-\mu^-$ and their experimental sensitivity at the LHCb. In both channels, we find that sensitivities on the branching fraction of the order $\lesssim 10^{-7} \ (10^{-8})$ might be accessible at the LHC run 2 (future LHC run 3), allowing us to set additional and complementary constraints on the parameter space associated with the mass and mixings of the Majorana neutrino. In particular, bounds can be obtained on the mixing $|V_{\mu N}|^2 \sim \mathcal{O}(10^{-5}-10^{-4})$ that are similar or better than the ones obtained from heavy meson $\Delta L=2$ decays: $D_{(s)}^- \to \pi^+\mu^-\mu^-$  and $B^- \to \pi^+\mu^-\mu^- \ (D^0\pi^+\mu^-\mu^-)$.
\end{abstract}

\maketitle
\bigskip

\section{Introduction}  \label{Intro}

Nowadays, the conclusive fact that neutrinos are massive particles (nonzero but very tiny masses $\lesssim 1$ eV) has been established by the accumulated evidence from neutrino oscillation experiments \cite{PDG}. This provides one of the strongest motivations for considering new physics scenarios beyond the Standard Model (SM) in order to get an explanation of the neutrino mass generation \cite{Reviews}. In addition, and closely linked to the previous fact, elucidating the nature of the neutrinos (Dirac or Majorana particles) is still an open issue in current particle physics \cite{Reviews}.

It is known that if neutrinos turn out to be of Majorana nature, their effects should manifiest through the observation of decay and production phenomena where the total lepton number $L$ is violated by two units ($\Delta L=2$). A clear signal of such a $\Delta L=2$ processes implies the production of same-sign dileptons in the final state. The most appealing of these lepton-number-violating (LNV) signatures is the neutrinoless double-beta ($0\nu\beta\beta$) nuclear decay, which has been regarded as the most sensitive way to looking for \cite{Rodejohann:2011,Gomez-Cadenas}.
Observation of this nuclear decay would establish the existence of LNV processes, thus implying that neutrinos are Majorana particles \cite{Rodejohann:2011,Vissani:2015,Gomez-Cadenas}. So far, the $0\nu\beta\beta$ decay seems to be a rather elusive process and has not yet been observed experimentally; there are only limits that have been obtained on their decay rates, allowing us to set bounds on the effective Majorana mass at the sub-eV level ($\sim 10^{-1}$ eV) \cite{Rodejohann:2011,Gomez-Cadenas}.

\begin{figure}[!t]
\includegraphics[scale=0.5]{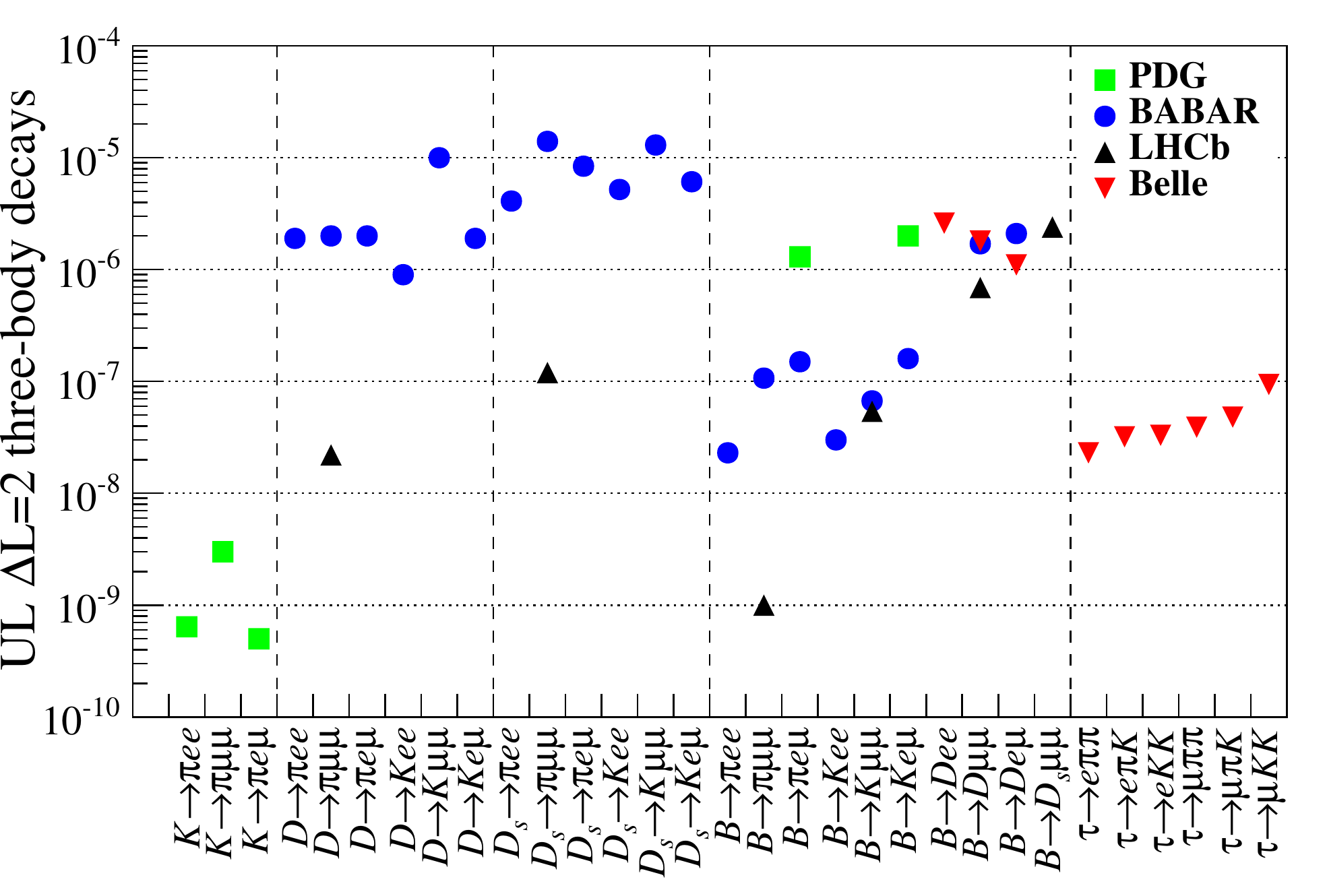}\hfill
\includegraphics[scale=0.5]{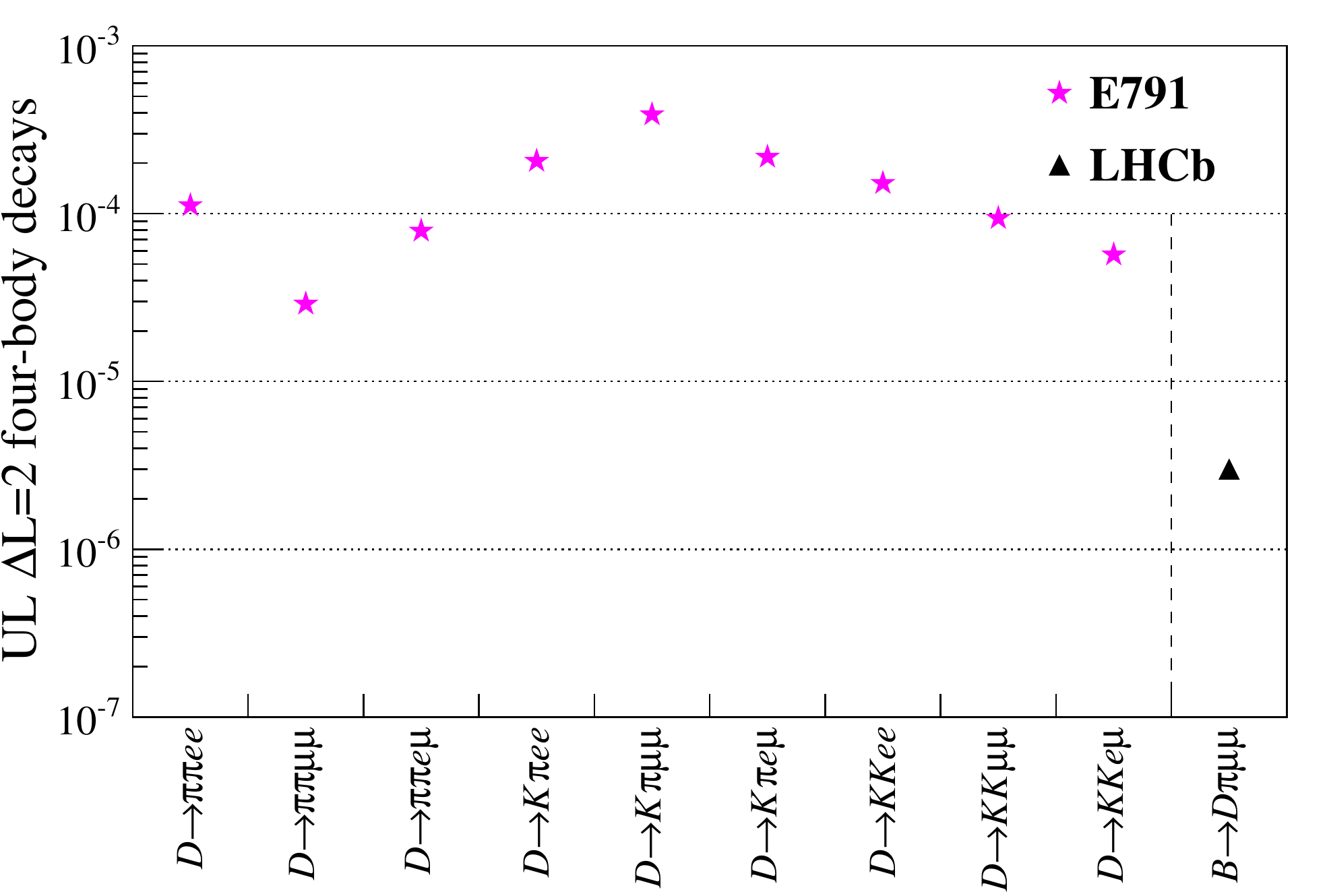}
\caption{\small Current experimental upper limits (UL) on branching ratios of three- (top) and four-body (bottom) $\Delta L=2$ decays \cite{Quintero:2012a}. Taken from PDG \cite{PDG} and BABAR \cite{BABAR,BABAR:2014}, LHCb \cite{LHCb:2012,LHCb:2013,LHCb:2014}, Belle \cite{Belle:2011,Belle:2013} and E791 \cite{E791} Collaborations. For simplicity we have not included the UL on $B^- \to V^+ \ell^- \ell^{\prime -}$, with $V=\rho,K^{\ast},D^{\ast}$, which are typically of the order $\lesssim 10^{-7}$ \cite{BABAR:2014}}
\label{fig:1} 
\end{figure}

Different $\Delta L=2$ processes at low and high energies have been proposed in the literature as alternative evidence to prove the Majorana nature of neutrinos (for a detailed list, see \cite{Atre:2009}).
Among all these possibilities, an interesting source of LNV processes is given by the exchange of a single Majorana neutrino ($N$) with a intermediate mass of the order of the heavy-flavors mass scale. This heavy neutrino can be produced on their mass-shell and the decay rates receive an enhancement due to the resonant effect \cite{Atre:2009}.
The effects of such a sterile heavy neutrino with masses around 100 MeV to a few GeV have been widely studied in $\Delta L=2$ three-body decays: $(K^-, D_{(s)}^-, B^-, B_c^-) \to M^+ \ell^- \ell^{\prime -}$  \cite{Atre:2009,Kovalenko:2000,Ali:2001,Atre:2005,Kovalenko:2005,Helo:2011,Cvetic:2010,
Zhang:2011,Bao:2013,Wang:2014} and
$\tau^- \to \ell^+ h^-h^{\prime -}$ \cite{Atre:2009,Helo:2011,Gribanov:2001}, with $ \ell^{(\prime)} = e, \mu$ and $M, h^{(\prime)}$ final state mesons.\footnote{The possibility of CP violation detection in $\Delta L=2$ decays of charged mesons has been recently explored in \cite{Cvetic:CP}.}
Moreover, $\Delta L=2$ four-body decays of $B$ and $D$ mesons: $\bar{B}^0 \to D^+\pi^+\ell^-\ell^-$, $\bar{D}^0 \to (\pi^+\pi^+,K^+\pi^+)\mu^{-}\mu^{-}$, and $B^{-} \to D^{0} \pi^{+}\mu^{-}\mu^{-}$ \cite{Quintero:2011,Quintero:2012b,Quintero:2013} (see also \cite{Dong:2013,Yuan:2013}), and $\tau$ lepton: $\tau^{-} \to \pi^+\nu_{\tau} \ell^-\ell^-$ \cite{Quintero:2012a,Quintero:2012b,Dib:2012} have been also studied as additional and complementary processes. 
Experimental efforts on searches of these $\Delta L=2$ decays have been reported by the Particle Data Group (PDG) and several experiments such BABAR, Belle, LHCb, and E791, but no evidence has been seen so far. Figure \ref{fig:1} summarizes the current upper limits on branching ratios of $\Delta L=2$ three- (top) and four-body (bottom) decays, respectively \cite{PDG,BABAR,BABAR:2014,LHCb:2012,LHCb:2013,LHCb:2014,Belle:2011,Belle:2013,E791}. The non-observation of these $\Delta L=2$ decays can be turned out into significant constraints on the parameters (mixings and masses) of a heavy Majorana neutrino \cite{Atre:2009,Helo:2011,Quintero:2012a,Quintero:2013}.

A well-motivated  and realistic scenario of heavy neutrinos with masses in the interval explored by those $\Delta L=2$ decays is motivated by the neutrino minimal SM ($\nu$MSM) \cite{Shaposhnikov:2005,Shaposhnikov:2013}. The Majorana mass spectrum of the $\nu$MSM contains  two almost degenerate Majorana neutrinos $\sim \mathcal{O}(1)$ GeV  and another lighter $\sim \mathcal{O}(10)$ keV, providing simultaneously an explanation of the generation of neutrino masses, dark matter, and the baryon asymmetry of the Universe (via leptogenesis) \cite{Shaposhnikov:2005,Shaposhnikov:2013,Drewes:2014}.
Thus, it is important studying the possible phenomenological implications of such a spectrum, especially the presence of a sterile neutrino with a mass $\mathcal{O}(1)$ GeV that can be produced and studied in different high-intensity frontier experiments \cite{Drewes:2014,Drewes:2013,Deppisch:2015,Drewes:2015}, rendering it to a falsifiable scenario.

In the present work, we extend the study of low-energy LNV processes exploring the $\Delta L=2$ decays of the $B_c$ meson, namely, three-body $B_c^- \to \pi^+\mu^-\mu^-$ [Fig. \ref{Fig:Bc_DeltaL2}(a)] and four-body $B_c^- \to J/\psi \pi^+\mu^-\mu^-$ [Fig. \ref{Fig:Bc_DeltaL2}(b)] decays, within the scenario provided by the production of an on-shell Majorana neutrino \cite{Atre:2009,Helo:2011}. The recent LHCb limits on the similar topology modes, $B^{-} \to \pi^{+}\mu^{-}\mu^{-}$ \cite{LHCb:2014} and $B^{-} \to D^{0} \pi^{+}\mu^{-}\mu^{-}$ \cite{LHCb:2012}, show the feasibility of exploring $\Delta L=2$ processes in $B^-$ meson decays. 
As the heaviest $b$-flavored meson, the $B_c^-$ meson opens the opportunity to many possible final states that enable us to explore signs of new physics. The search for same-sign di-leptonic signatures at $B_c^-$ meson decays can also be explore at LHCb, particularly in the same-sign di-muon  ($\mu^-\mu^-$) channel, due to the relatively high muon reconstruction system. In addition, an  appealing characteristic is that annihilation [Fig. \ref{Fig:Bc_DeltaL2}(a)] and spectator [Fig. \ref{Fig:Bc_DeltaL2}(b)] diagrams are not very much suppressed by CKM factors.
As we will show, this fact enables us to get stronger bounds on the parameter space associated with the mass and mixings of the heavy Majorana neutrino, as pointed out in \cite{Quintero:2012a}.




In the following, in Sec. \ref{Sec:Bc_DeltaL2} we study the $\Delta L = 2$ decays of the $B_c$ meson: $B_c^- \to \pi^+\mu^-\mu^-$ and $B_c^- \to J/\psi \pi^+\mu^-\mu^-$. The experimental sensitivity at the LCHb and constraints on the $(m_N,|V_{\mu N}|^2)$ plane are also studied. The comparison with different search strategies is briefly discussed in Sec. \ref{comparison}. Our conclusions are presented in Sec. \ref{conclusions}.

\section{$\Delta L = 2$ decays of $B_c$ meson}  \label{Sec:Bc_DeltaL2} 

\begin{figure}[!t]
\centering
\includegraphics[scale=0.50]{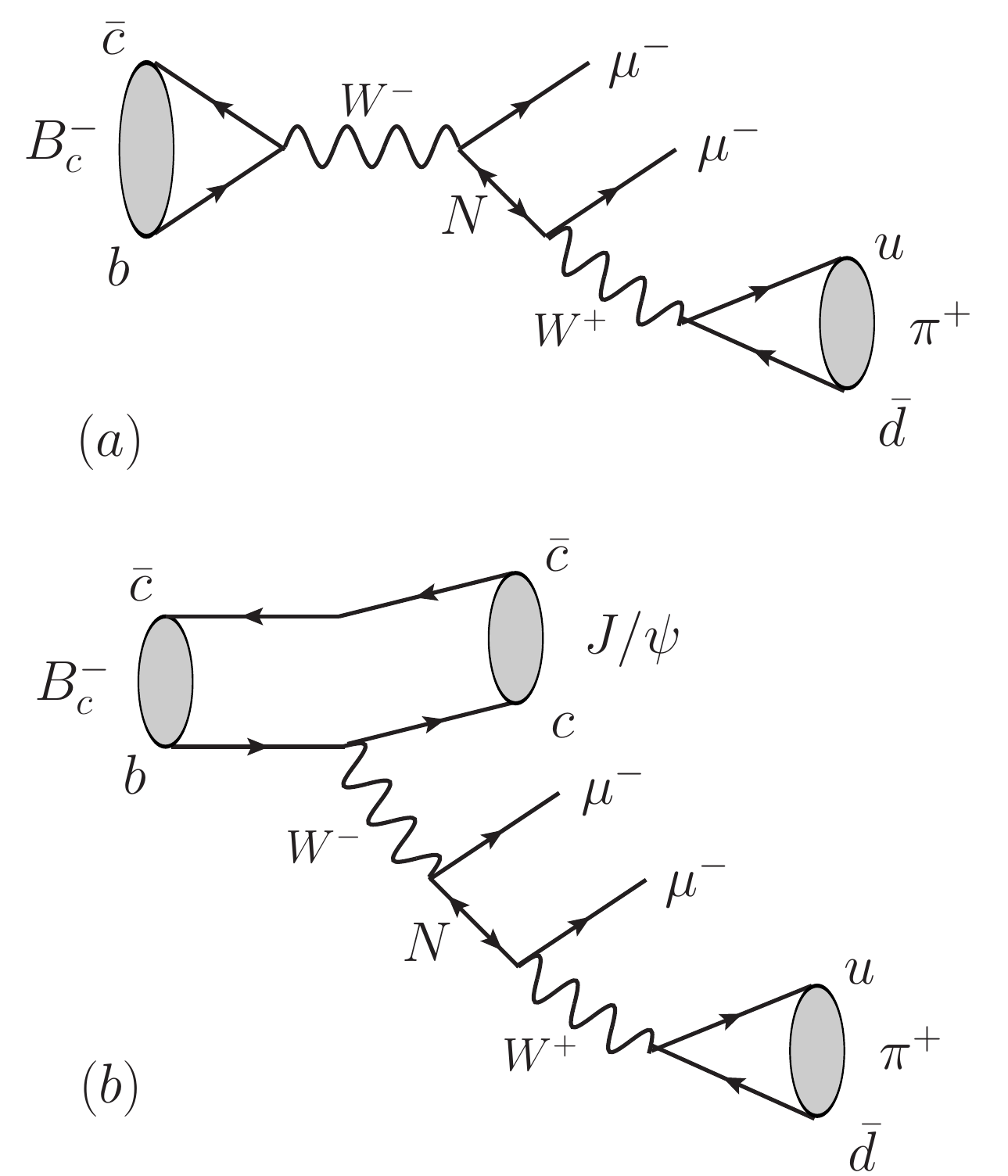}
\caption{\small $\Delta L = 2$ decays of $B_c$ meson mediated by a Majorana neutrino $N$: (a) three-body $B_c^- \to \pi^+\mu^-\mu^-$ and (b) four-body $B_c^- \to J/\psi \pi^+\mu^-\mu^-$ channels.}
\label{Fig:Bc_DeltaL2} 
\end{figure}

The $\Delta L = 2$ decays of the $B_c$ meson can occur via the exchange of a Majorana neutrino with a kinematically allowed mass. Although different same-sign di-leptons ($\ell^-\ell^{\prime -}$ with $\ell^{(\prime)} = e, \mu, \tau$) are allowed, in this work we focus on the $\mu^-\mu^-$ channels in the three-body $B_c^- \to \pi^+\mu^-\mu^-$ and four-body $B_c^- \to J/\psi \pi^+\mu^-\mu^-$ decays as shown in Figs. \ref{Fig:Bc_DeltaL2}(a) and \ref{Fig:Bc_DeltaL2}(b), respectively. As in previous studies \cite{Atre:2009,Helo:2011,Zhang:2011,Cvetic:2010,Bao:2013}, we assume that only one heavy neutrino $N$, with a mass such that it can be produced resonantly (on-shell) in $B_c$ decays, dominates the decay amplitude.
Although the channel $B_c^- \to \pi^+\mu^-\mu^-$ has been previously considered in Refs. \cite{Cvetic:2010,Bao:2013}, we provide a reanalysis of this and its experimental signal at the LHCb, as well as an analysis of the channel $B_c^- \to J/\psi \pi^+\mu^-\mu^-$.

\subsection{$B_c^- \to \pi^+\mu^-\mu^-$}  \label{Bc3body}

Let us now explore the $\Delta L = 2$ four-body channel $B_c^- \to J/\psi \pi^+\mu^-\mu^-$. The diagram that contributes to this process is shown in Fig. \ref{Fig:Bc_DeltaL2}(b), which receives the effect of a kinematically accessible Majorana neutrino with a mass in the range 0.25 GeV $\leq m_N \leq$ 3.18 GeV. 

The $B_c^- \to \pi^+\mu^-\mu^-$ channel is induced by diagram \ref{Fig:Bc_DeltaL2}(a), and it can be enhanced for Majorana neutrino masses in the range 0.25 GeV $\leq m_N \leq$ 6.16 GeV. In this mass range, the total decay width of the intermediate Majorana neutrino $N$ $(\Gamma_N)$ is much smaller than its mass, $\Gamma_N \ll m_N$ \cite{Atre:2009}, so the narrow width approximation (NWA) is valid. This allows us to consider the Majorana neutrino as a particle that is produced on its mass shell through the leptonic decay $B_c^- \to \mu^-N$, followed by the subsequent decay $N \to \mu^-\pi^+$.  

The decay rate of the process $B_c^- \to \pi^+\mu^-\mu^-$ is then split into two subprocesses
\beq
\Gamma(B_c^- \to \pi^+\mu^-\mu^-) = \Gamma(B_c^- \to \mu^- N) \times \Gamma(N \to \pi^+\mu^-)
/\Gamma_N ,
\eeq

\noindent  with the decay widths of $B_c^- \to \mu^-N$ and $N \to \pi^+\mu^-$ given by the expressions \cite{Atre:2009}
\beq 
\Gamma(B_c^- \to \mu^- N)
= \dfrac{G_F^2}{8 \pi}|V_{cb}^{\text{CKM}}|^2 |V_{\mu N}|^2 f_{B_c}^2  \ m_{B_c} m_N^2 \Big(1- \dfrac{m_N^2}{m_{B_c}^2} \Big)^2 , \label{BctomuN}
\eeq
\beq 
\Gamma(N \to \mu^-\pi^+)
= \dfrac{G_F^2}{16 \pi}|V_{ud}^{\text{CKM}}|^2 |V_{\mu N}|^2 f_\pi^2  \ m_N^3 \Big(1- \dfrac{m_\pi^2}{m_N^2} \Big)^2 , \label{Ntopimu}
\eeq

\noindent respectively, where $|V_{cb}^{\rm{CKM}}| = 41.1 \times 10^{-3}$ and $|V_{ud}^{\text{CKM}}|= 0.97425$ are the Cabibbo-Kobayashi-Maskawa (CKM) quark mixing matrix elements \cite{PDG}, $G_F$ is the Fermi constant, and $f_{\pi}= 130.2(1.4)$ MeV \cite{Rosner:2015} and $f_{B_c}= 434(15)$ MeV \cite{HPQCD} are the decay constants of the mesons involved. For simplicity in \eqref{BctomuN} and \eqref{Ntopimu}, the muon mass has been neglected. The coupling of the heavy neutrino $N$ to the charged current of lepton flavor $\mu$ is characterized by the quantity $V_{\mu N}$ \cite{Atre:2009}. 

The full decay width of the Majorana neutrino $\Gamma_N$ is obtained by adding up the contributions of all the neutrino decay channels that can be opened at the mass $m_N$ \cite{Atre:2009}. The dominant decay modes of the neutrino in the range of masses relevant for the $\Delta L=2$ channel under consideration are the following: semileptonic two-body and leptonic three-body decays \cite{Atre:2009}.

\subsection{$B_c^- \to J/\psi \pi^+\mu^-\mu^-$}   \label{Bc4body}

Let us now explore the $\Delta L = 2$ four-body channel $B_c^- \to J/\psi \pi^+\mu^-\mu^-$. The diagram that contributes to this process is shown in Fig. \ref{Fig:Bc_DeltaL2}(b), which receives the effect of a kinematically accessible Majorana neutrino with a mass in the range 0.25 GeV $\leq m_N \leq$ 3.18 GeV. 
This sort of four-body channels was considered the first time in $B$ and $D$ decays \cite{Quintero:2011,Quintero:2012a,Quintero:2013} (see also \cite{Dong:2013,Yuan:2013}), but has thus far have not been considered in the literature on $B_c$ decays. The dynamics associated with this type of four-body decays is completely different from that considered in Sec. \ref{Bc3body} (and any $\Delta L=2$ three-body decays). Thus, it offers an additional LNV signal to search for and test the heavy neutrino sector.

By the same NWA arguments previously considered in Sec. \ref{Bc3body}, we consider the Majorana neutrino as an intermediate particle that is produced on it mass shell in the semileptonic decay $B_c^- \to J/\psi \mu^-N$, followed by the subsequent decay $N \to \mu^-\pi^+$. In the on-shell factorization, the channel $B_c^- \to J/\psi \pi^+\mu^-\mu^-$ is written as 
\beq
\Gamma(B_c^- \to J/\psi \pi^+\mu^-\mu^-) = \Gamma(B_c^- \to J/\psi \mu^- N) \times \Gamma(N \to \mu^-\pi^+)/\Gamma_N ,
\eeq

\noindent where the decay width of the subprocess $N \to \pi^+\mu^-$ is given by Eq. \eqref{Ntopimu}. In the charged lepton massless limit, the decay width of $B_c^- \to J/\psi \mu^-N$ is given by the expression (see Appendix \ref{appA} for details)
\beq 
\Gamma(B_c^- \to J/\psi \mu^- N)
= \dfrac{G_F^2}{32 (2\pi)^3 m_{B_c}^3} |V_{cb}^{\text{CKM}}|^2 |V_{\mu N}|^2 \int_{m_N^2}^{(m_{B_c} - m_{J/\psi})^2} dt \ \mathcal{A}(t) , \label{BctopimuN}
\eeq

\noindent where $\mathcal{A}(t)$ is a function of the squared transfer momentum $t=(p_{B_c} - p_{J/\psi})^2$ [see Eq. \eqref{functionA}]. In the heavy quark spin symmetry the hadronic transition $B_c \to J/\psi$ can be parametrized in terms of a single form factor $\Delta^{B_c \to J/\psi}$ \cite{HQSS,Colangelo:2000}. Based on the framework of a QCD relativistic potential model \cite{Colangelo:2000}, the form factor can be represented by the three-parameter formula
\beq \label{Delta}
\Delta^{B_c \to J/\psi}(y) = \Delta(0) [1- \rho^2(y-1) + c(y-1)^2],
\eeq

\noindent in terms of the value at zero recoil $\Delta(0)=0.94$, the slope $\rho^2 = 2.9$ and the
curvature $c=3$ \cite{Colangelo:2000}. The variable $y$ is related to $t$ by the relation $y=(m_{B_c}^2 + m_{J/\psi}^2 - t)/ 2m_{B_c}m_{J/\psi}$ \cite{Colangelo:2000}\footnote{In ensuing numerical evaluation we will use the theoretical predictions provided by this model because its results are in
good agreement with recent measurements of the hadronic modes $B_c^+ \to J/\psi K^+$ and $B_c^+ \to J/\psi D_s^{(\ast) +}$ reported by the LHCb Collaboration \cite{LHCb:Bc}.}. 

\subsection{Experimental sensitivity at the LCHb and constraints on $(m_N,|V_{\mu N}|^2)$ plane}  

The number of expected events in the LHCb experiment in terms of the integrated luminosity has the form
\beq \label{Nexp}
N_{\rm exp} = \sigma(B_c)\mathcal{B}(B_c \to \Delta L=2) \epsilon_D \mathcal{L}_{\rm int} ,
\eeq

\noindent where  $\sigma(B_c)$ is the production cross-section of $B_c$ mesons inside the LHC geometrical acceptance, $\mathcal{B}(B_c \to \Delta L=2)$ corresponds to the estimated branching
fraction of the $\Delta L=2$ process in consideration, $\epsilon_D$ is the detection efficiency of the LHCb detector involving reconstruction, selection, trigger and particle misidentification efficiencies, and $\mathcal{L}_{\rm int}$ is the integrated luminosity.

The cross section $\sigma(B_c)$ has never been measured. With respect to the cross section of other $b$ mesons, the production of $B_c$ mesons is suppressed by a hard production of the additional $c$-quark and for the small probability of the $\bar{b}c$ bound state formation. Ref. \cite{Likhoded} states that given above constrains, a good approach is to consider $\sigma(B_c)/\sigma(B_u + B_d + B_s) \sim 10^{-3}$. The LHCb experiment has measured $\sigma(B_{u,d})$ and $\sigma(B_s)$, at a center of mass energy of 7 TeV and inside the LHCb geometrical acceptance, to be $\sigma(B_{u,d,s} , 7 \ {\rm TeV})$ = 87.5 $\pm$ 6.7 $\mu$b, where the error is given by statistical and systematic uncertainties \cite{LHCb:sigma}. Assuming that the $b$-quark cross sections scale with energy
in the same way as the $c$-charm cross sections and based on the LHCb measurements
of the prompt charm cross sections at 7 and 13 TeV \cite{LHCb:c_prompt} and assuming a linear scaling of the production cross section with the center of mass energy, we can roughly estimate within the LHCb acceptance $\sigma(B_c , 7 \ {\rm TeV}) \sim 80$ nb, $\sigma(B_c , 8 \ {\rm TeV}) \sim 90$ nb and $\sigma(B_c , 14 \ {\rm TeV}) \sim 160$ nb, where a conservative approach for the uncertainty on this quantity is a relative 25\%.

The LHCb experiment performance during LHC run 1 can be found at \cite{LHCb:performance}. During run 1, the LHCb detector collected at $\sqrt{s} = 7$ TeV an integrated luminosity of 1 fb${}^{-1}$ and 2 fb${}^{-1}$ at $\sqrt{s} = 8$ TeV. The plan for LHC run 2 is to collect additional 5 fb${}^{-1}$ at the LHC nominal construction energy of center of mass of 14 TeV. Already some work have been developed for future LHCb upgrade, LHC run 3, where integrated luminosities of the order of 40 fb${}^{-1}$ are expected.

Precise computation of the detection efficiency requires fully simulated decay specific Monte Carlo samples, reconstructed in the same manner as real data and with a simulation of the full detector. However a rough estimation can be done with detection efficiencies already reported by LHCb experiment in the study of some $B_c$ decays. The efficiency will depend on the topological and kinematical signatures of the decay. In Ref. \cite{LHCb:Bcproduction}, the study of the $B_c^+\to J/\psi\pi^+$ is performed, finding a total signal efficiency of 6\% using a multivariate tight selection. In the LNV mode $B_c^-\to \pi^+\mu^-\mu^-$, we have same final states particle and, thus, it is natural to assume the same reconstruction efficiency, while in $B_c^-\to J/\psi (\to \mu^+\mu^-)\pi^+\mu^-\mu^-$ it shares part of the mentioned decay with two additional muon tracks. The addition of more tracks reduces the detection efficiency; thus, 6\% represents an upper limit for our case. Muon tracks traverse the full LHCb detector and are reconstructed using information of all the subdetectors, keeping a relatively high reconstruction efficiency of about 90\% for single tracks; therefore, assuming $\epsilon_D=5\%$ for the second decay mode is a conservative value for detector and reconstruction conditions for LHCb during LHC run 1. In the less optimistic case, this efficiency will not grow for LHC run 2 and upgrade periods. This will not be the case given the great effort of the LHCb Collaboration to improve the detector itself and the reconstruction algorithms, but we will keep this value for further computations, with a relative 25\% uncertainty.

\begin{figure}[!b]
\centering
\includegraphics[scale=0.6]{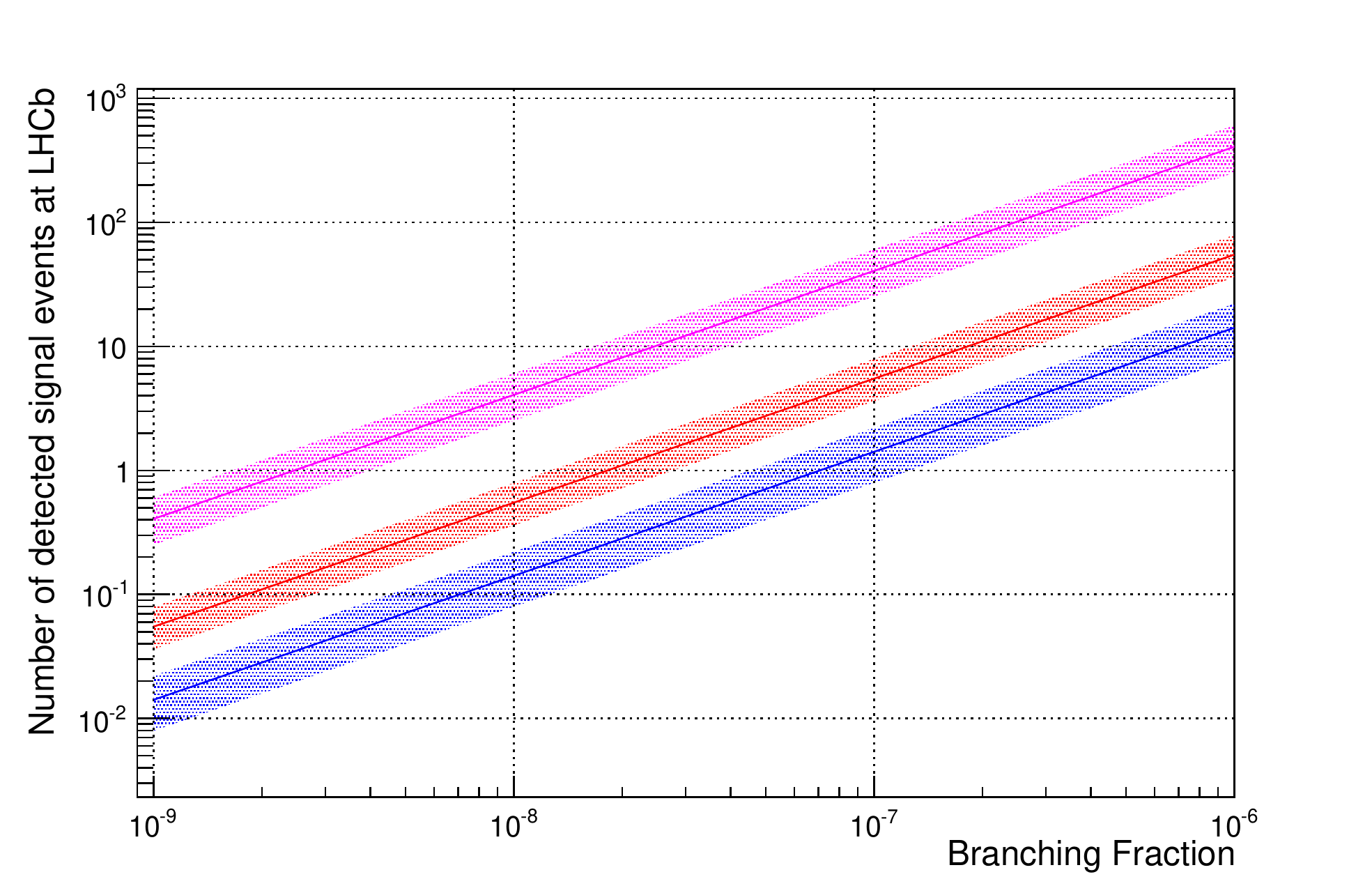}
\caption{\small Number of expected events of the $\Delta L=2$ process $B_c^-\to \pi^+\mu^-\mu^-$ and $B_c^-\to J/\psi \pi^+\mu^-\mu^-$ to
be observed in the LHCb experiment as a function of the branching ratio, for the different integrated luminosity collected: Blue LHC run 1 (3 fb${}^{-1}$), red LHC run 2 (5 fb${}^{-1}$), and purple LHC run 3 (40 fb${}^{-1}$). Filled area shows the uncertainty in the computation.}
\label{Nevents_vs_BF}
\end{figure}

We assume a value of detection efficiency of $\epsilon_D=5.5 \%$ for both $B_c^-\to \pi^+\mu^-\mu^-$ and $B_c^-\to J/\psi \pi^+\mu^-\mu^-$, and taking the above assumptions on cross section, Fig. \ref{Nevents_vs_BF} shows the number of expected events to be
observed at the LHCb experiment as a function of the branching fraction, to different integrated luminosity collected by LHCb. The figure shows that, at the end of LHC run 1, only a couple of events could have been detected, at the end of LHC run 2, only 5 events in the best scenario, and at the end of LHC run 3, the LHCb experiment could detect up to 35 signal events of the given $\Delta L=2$ process.
For the LHC run 2 and LHC run 3, this can be translated into expected sensitivities on the branching fraction of the order $\lesssim 10^{-7}$ and $\lesssim 10^{-8}$, respectively. In the following, we will take those values of branching fractions as the most conservative ones. 

\begin{figure}[!b]
\centering
\includegraphics[scale=0.6]{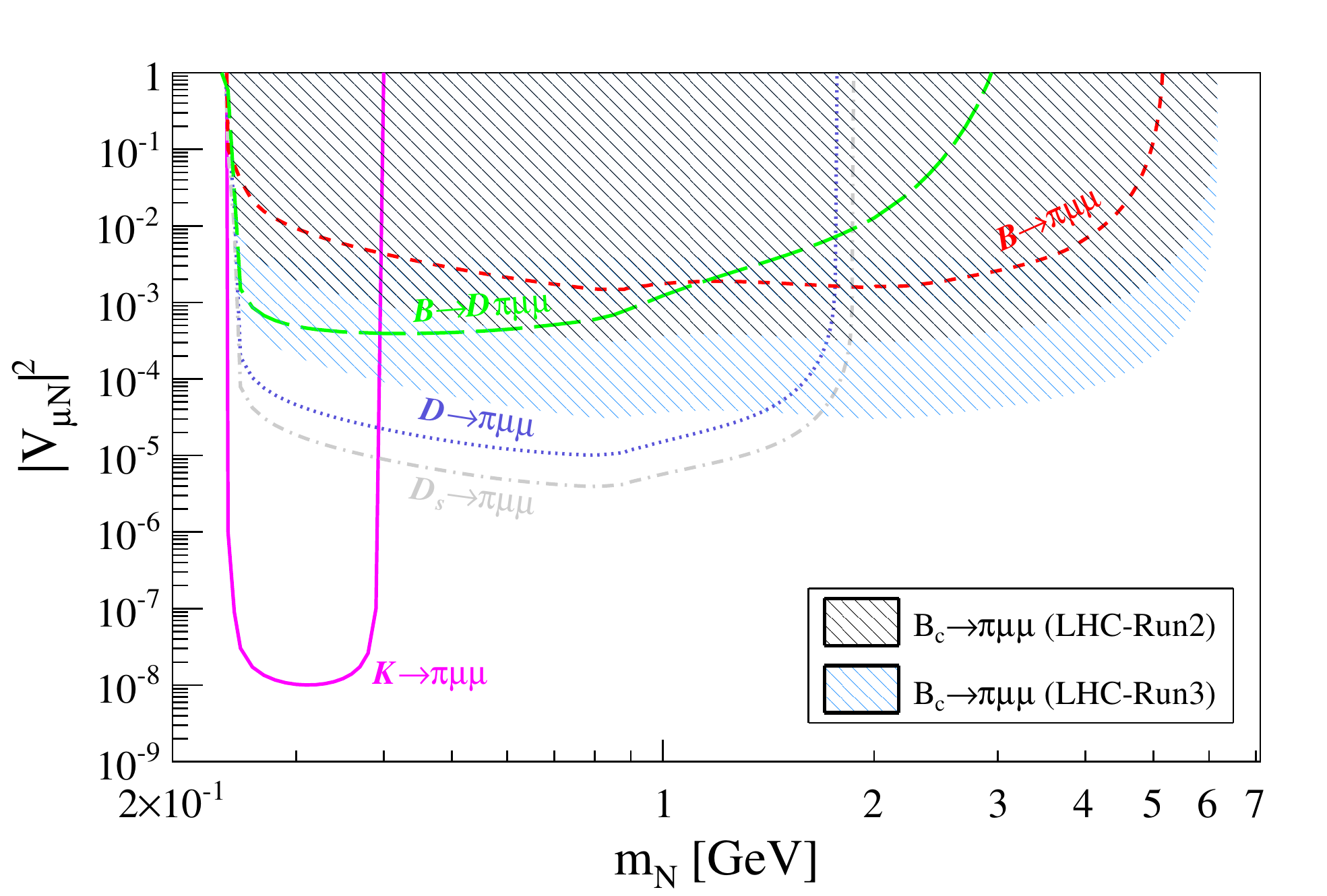}
\caption{\small Exclusion regions on $(m_N, |V_{\mu N}|^{2})$ plane, from LNV di-muon modes. The black (blue) region represents the constraint obtained from the search on $B_c^- \to \pi^+\mu^-\mu^-$ at LHCb for LHC run 2 (LHC run 3). Updated constraints provided by $(K^-, D_{(s)}^{-}, B^{-}) \to \pi^{+}\mu^{-}\mu^{-}$ and $B^- \to D^0\pi^+\mu^-\mu^-$ are also included for comparison.}
\label{VmuNBc3body}
\end{figure}

In order to illustrate the constraints that can be achieved from the experimental searches on $B_c^- \to \pi^+\mu^-\mu^-$, in Fig. \ref{VmuNBc3body} the black (blue) area represents the excluded regions on $|V_{\mu N}|^{2}$ as a function of $m_N$ obtained by taking branching fraction of the order $\mathcal{B}(B_c^- \to \pi^+\mu^-\mu^-) \lesssim 10^{-7} \ (10^{-8})$ at the LHCb for LHC run 2 (LHC run 3).
For comparison, we also display the updated exclusion limits obtained from searches on $\Delta L=2$ channels $(K^-, D_{(s)}^{-}, B^{-}) \to \pi^{+}\mu^{-}\mu^{-}$  \cite{PDG,LHCb:2013,LHCb:2014}\footnote{It is worth noticing that we have not included the limit obtained by LHCb because it is overestimated and does not give the expected behavior as demanded by CKM factors and neutrino mass \cite{LHCb:2014}.} and $B^- \to D^0\pi^+\mu^-\mu^-$ \cite{Quintero:2013,LHCb:2012}. 
For neutrino mass values below 2 GeV, we can see that the most restrictive constraint is given by $K^- \to \pi^+ \mu^-\mu^-$ which can reach $|V_{\mu N}|^2\sim \mathcal{O}(10^{-8})$ but only for a very narrow range $[0.25, 0.38]$ GeV of Majorana neutrino masses. 
Followed in sensitivity by $D_{s}^{-} \to \pi^{+}\mu^{-}\mu^{-}$ and $D^{-} \to \pi^{+}\mu^{-}\mu^{-}$, $|V_{\mu N}|^2 \sim \mathcal{O}(10^{-6} - 10^{-5})$, 
covering a wider window mass  [0.24,1.86] GeV and [0.24,1.76] GeV, respectively.
In the region where these overlap, both LHC run 2 and LHC run 3, the Cabibbo-allowed channel $B_c^- \to \pi^+\mu^-\mu^-$ could almost exclude the whole region covered by $B^- \to D^0\pi^+\mu^-\mu^-$ (long-dashed line) and $B^- \to \pi^+\mu^-\mu^-$ (short-dashed line), despite the fact that the upper limit on the latter Cabibbo-suppressed channel is of order $10^{-9}$ \cite{LHCb:2014}.  On the other hand, for $m_N \gtrsim 2$ GeV, the channel $B_c^- \to \pi^+\mu^-\mu^-$ would reach a mass region up to $\sim 6$ GeV not yet covered for any other LNV search.

For the case $B_c^- \to J/\psi\pi^+\mu^-\mu^-$, with an expected sensitivity at the LCHb of $\mathcal{B}(B_c^- \to J/\psi\pi^+\mu^-\mu^-) \lesssim 10^{-7} \ (10^{-8})$ for LHC run 2 (LHC run 3), in Fig. \ref{VmuNBc3body} we show the exclusion curves on $(m_N,|V_{\mu N}|^2)$ represented by the black (blue) region.
Similar remarks as the ones previously discussed (see Fig. \ref{VmuNBc3body}) can be med for neutrino mass values below $\sim 2$ GeV. In this case, in the region where these overlap, the Cabibbo-allowed four-body channel under consideration is able to exclude a larger region of $|V_{\mu N}|^{2}$ than the channel $B^- \to D^0\pi^+\mu^-\mu^-$ (long-dashed line) and Cabibbo-suppressed channel $B^- \to \pi^+\mu^-\mu^-$ (short-dashed line). Again, this is despite the fact that the upper limit on the latter is of order $10^{-9}$ \cite{LHCb:2014}. At the end of the LHC run 3 the search on $B_c^- \to J/\psi\pi^+\mu^-\mu^-$ would almost exclude all the region covered by $D_{(s)}^- \to \pi^+\mu^-\mu^-$.	

\begin{figure}[!t]
\centering
\includegraphics[scale=0.6]{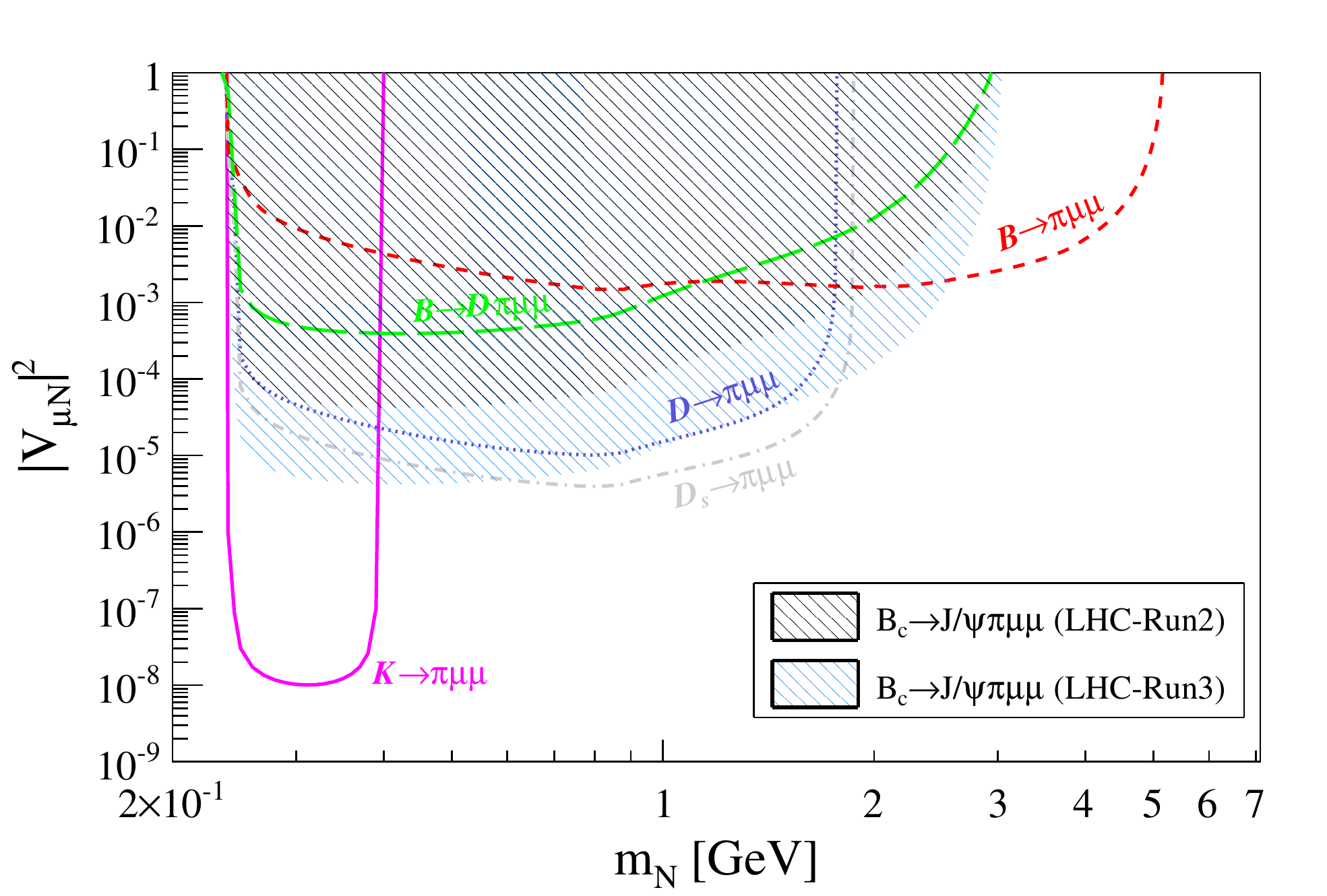}
\caption{\small Same description as in Fig. \ref{VmuNBc3body}, but the black (blue) region represents the constraint obtained from the search on $B_c^- \to J/\psi \pi^+\mu^-\mu^-$ at LHCb for LHC run 2 (LHC run 3).}
\label{VmuNBc}
\end{figure}

We end this section by mentioning that, to this point, the intermediate heavy Majorana neutrino has been treated as a real particle which propagates before decaying. Most of the the on-shell neutrinos produced in the decays $B_c^- \to \mu^- N$ and $B_c^- \to J/\psi \mu^- N$, are expected to live a long enough time to travel through the detector and decay ($N \to  \pi^+\mu^-$) far from the interaction region.
This effect is given by the acceptance factor ($P_N$), which represents the probability for the on-shell neutrino $N$ decay products to be inside the detector acceptance \cite{Atre:2009,Helo:2011,Dib:2014}. The reconstruction efficiency $\epsilon_D$ will depend on this acceptance factor as well, suppressing the number of detected decays. A precise estimation will require knowledge of the production and decay vertex, which can be adequately included by LHCb during the data analysis.

\section{Comparison with other experimental searches}  \label{comparison}

It is important to comment that for the range of sterile neutrino masses relevant to this work, 0.2 GeV $\lesssim m_N \lesssim$ 6 GeV, different search strategies have been used to get constraints on the mixing element $|V_{\mu N}|$ (for a recent review on theoretical and experimental status see \cite{Atre:2009,Deppisch:2015,Drewes:2013,Drewes:2015,deGouvea:2015} and references therein). The lack of experimental evidence of searches of peaks in the muon spectrum of leptonic $K^{\pm}$ decays (PS191, E949) and searches through specific visible channels of heavy neutrino decays produced in beam dump experiments (NA3, CHARM, BEBC, FMMF, NuTeV) allows us to put constraints on $|V_{\mu N}|^2 \sim \mathcal{O}(10^{-8} - 10^{-6})$  for masses of the  sterile neutrino ranging from 0.2 to 2 GeV \cite{Atre:2009,Deppisch:2015,Drewes:2013,Drewes:2015,deGouvea:2015}. Moreover,
in the mass range [0.5,5] GeV searches of heavy neutrinos have been performed by Belle Collaboration (data sample of 772 million $B\bar{B}$ pairs) using the inclusive decay mode $B \to X \ell N$ followed by $N \to \ell\pi$ (with $\ell = e, \mu$), reaching a sensitivity of $|V_{\mu N}|^2 \sim \mathcal{O}(10^{-5})$ \cite{Belle:N}\footnote{In addition, it is expected that the recently proposed high-intensity beam dump experiment SHiP \cite{SHiP}, can significantly improve those bounds through search of such a heavy neutrinos in the leptonic and semihadronic decays of $D_s$ mesons \cite{Deppisch:2015,Helo:2015}.}. In comparison, the current constraints obtained from direct searches of LNV ($\Delta L=2$) decays, including the ones considered in this work, are certainly less restrictive except for the $K^- \to \pi^+ \mu^-\mu^-$ channel (see Figs. \ref{VmuNBc3body} and \ref{VmuNBc}). 

In the mass region 2 GeV $< m_N <$ 5 GeV, the search of $B_c^- \to \pi^+\mu^-\mu^-$ would exclude similar regions as the ones obtained from DELPHI using the possible production of heavy neutrinos in the $Z$-boson decay $Z \to \nu N$ ($|V_{\mu N}|^2 \leq 10^{-4}$) \cite{LEP} and Belle bounds  \cite{Belle:N}. For masses $m_N < m_W$, the possibility of a heavy neutrino produced in $W$ and Higgs boson decays has been studied as well \cite{Helo:2014,Izaguirre:2015,Gago:2015,Dib:2015,BhupalDev:2012}.

\section{Conclusions}  \label{conclusions}
 
A Majorana neutrino with a mass around $\sim$ 0.2 GeV to a few GeV can be produced on its mass shell
at high-intensity frontier experiments and its effect is manifested through lepton-number-violating (LNV) processes, where the total lepton number $L$ is violated by two units ($\Delta L=2$). Within this Majorana neutrino scenario, in the present work we have extended the previous studies of low-energy LNV processes exploring the $\Delta L=2$ decays of $B_c$ meson: three-body $B_c^- \to \pi^+\mu^-\mu^-$ and four-body $B_c^- \to J/\psi \pi^+\mu^-\mu^-$, which are not very much suppressed by CKM factors.  
Due to the relatively high muon reconstruction system, we have paid attention on these same-sign di-muon channels and explored their experimental sensitivity at the LHCb. As conservative values, in both channels we have found that for a integrated luminosity collected of 5 fb${}^{-1}$ (LHC run 2) and 40 fb${}^{-1}$ (LHC run 3), one would expect sensitivities on the branching fraction of the order $\lesssim 10^{-7}$ and $\lesssim 10^{-8}$, respectively. With such sensitivities, in the best case, we can get additional and complementary constraints on the mixing parameter $|V_{\mu N}|^2 \sim \mathcal{O}(10^{-5} - 10^{-4})$ for neutrino masses in the range 0.2 $ \lesssim m_N \lesssim$ 6 GeV. These bounds are similar to or better than the ones obtained from heavy meson $\Delta L=2$ decays: $D_{(s)}^- \to \pi^+\mu^-\mu^-$  and $B^- \to \pi^+\mu^-\mu^- \ (D^0\pi^+\mu^-\mu^-)$.




\medskip


\textit{Note added.}-- After the completion of the present work and while we were preparing the draft, we became aware of version 2 of \cite{Sinha:2016}, where the $\Delta L=2$ decays of the $B_c$ meson under consideration in this paper are also studied. In \cite{Sinha:2016} different same-sign di-lepton channels $B_c^- \to \bar{B}_s^0 \pi^+\ell^-\ell^{\prime -}$, $B_c^- \to J/\psi \pi^+\ell^-\ell^{\prime -}$, and $B_c^- \to \pi^+ \ell^-\ell^{\prime -}$ have been considered, with $\ell^{(\prime)} = e, \mu, \tau$. The first channel, although CKM favoured, is also phase-space suppressed, allowing us to explore a narrow range of neutrino masses $[0.1,0.9]$ GeV. 
We have only focused on the second and third $\mu^-\mu^-$ channels, and aside from the different approach used to describe the form factors associated with the $B_c \to J/\psi$ transition (the authors of \cite{Sinha:2016} used QCD sum rules to estimate those form factors), in general, our conclusions   are in agreement with theirs  \cite{Sinha:2016}.
A more elaborate analysis of the experimental sensitivity at the LHCb of these $\Delta L=2$ modes is provided in the present study. Moreover, a more detailed comparison with the exclusion limits obtained from searches on $\Delta L=2$ channels $(K^-, D_{(s)}^{-}, B^{-}) \to \pi^{+}\mu^{-}\mu^{-}$ and $B^- \to D^0\pi^+\mu^-\mu^-$ is provided in our study.

\begin{acknowledgments}
The authors N. Q. and C. E. V. would like to thank Comit\'{e} Central de Investigaciones of the University of Tolima for financial support under Project No. 330115. ​We are grateful to Gabriel L\'{o}pez Castro for reading the manuscript and helpful suggestions.
\end{acknowledgments}


\appendix

\section{Semileptonic decay $B_c^- \to J/\psi \mu^-N$}  \label{appA}

The decay amplitude of the semileptonic subprocess $B_c^-(p_{B_c}) \to J/\psi(p_{J/\psi}) \mu^-(p_\mu)N(p_N)$ is given by the expression 
\beq 
{\cal M}(B_c^- \to J/\psi \mu^-N)
= \dfrac{G_F}{\sqrt{2}} V_{cb}^{\text{CKM}} V_{\mu N} \langle J/\psi |\bar{b}\gamma^{\alpha}\gamma_5 c|B_c \rangle \ [\bar{u}(p_\mu) \gamma_{\alpha} (1-\gamma_{5}) v(p_N)], \label{AmpliA}
\eeq

\noindent where $G_F$ is the Fermi constant, $V_{cb}^{\text{CKM}}$ is the CKM matrix element, and $V_{\mu N}$ denotes the mixing element of a muon with the heavy neutrino (sterile) $N$ in the charged current interaction \cite{Atre:2009}. In the heavy quark spin symmetry the hadronic current can be parametrized as \cite{HQSS,Colangelo:2000}
\beq 
\langle J/\psi |\bar{b}\gamma^{\alpha}\gamma_5 c|B_c \rangle = \sqrt{4 m_{B_c}m_{J/\psi}} \ \Delta^{B_c \to J/\psi} \epsilon^{\alpha *},
\eeq

\noindent relating to a single form factor $\Delta^{B_c \to J/\psi}$, with $\epsilon^{\alpha}$ the $J/\psi$ polarization vector. Based on the framework of a QCD relativistic potential
model \cite{Colangelo:2000}, the form factor can be represented by the three-parameter formula Eq. \eqref{Delta}. 

The decay width is parametrized in terms of the three-body phase space \cite{PDG},
\begin{equation}
\Gamma(B_c^- \to J/\psi \mu^-N) = \dfrac{1}{32(2\pi)^{3} m_{B_c}^{3}}  \int_{t^{-}}^{t^{+}}  dt \int_{s^{-}}^{s^{+}} ds \ |\overline{\mathcal{M}}|^{2},
\end{equation}

\noindent where $|\overline{\mathcal{M}}|^{2}$ is the spin-averaged squared amplitude and $s=(p_{J/\psi} + p_N)^{2}$ and $t=(p_{B_c} - p_{J/\psi})^2$ are kinematical variables. In the charged lepton massless limit, the integration limits are given by $t^{-} = m_{N}^{2}$, $t^{+} = (m_{B_c} - m_{J/\psi})^{2}$ and
\begin{eqnarray}
s^{\pm}(t) =&& m_{B_c}^2 + m_N^2 - \dfrac{1}{2 t} \Big[ (t + m_{B_c}^2 - m_{J/\psi}^2)(t + m_N^2) \nonumber\\
&& \mp \ \lambda_t^{1/2} (m_N^2 - t) \ \Big],
\end{eqnarray}

\noindent where $\lambda_t \equiv \lambda(t,m_{B_c}^2,m_{J/\psi}^2)$,  with $\lambda(x,y,z)=x^{2}+y^{2}+z^{2}-2xy-2xz-2yz$ the usual kinematic K\"{a}llen function. After integration over kinematical variable $s$, the decay width can be written as
\beq 
\Gamma(B_c^- \to J/\psi \mu^- N)
= \dfrac{G_F^2}{32 (2\pi)^3 m_{B_c}^3} |V_{cb}^{\text{CKM}}|^2 |V_{\mu N}|^2 \int_{m_N^2}^{(m_{B_c} - m_{J/\psi})^2} dt \  \mathcal{A}(t) , \label{BctopimuN}
\eeq

\noindent with 
\bea \label{functionA}
\mathcal{A}(t) &=& \dfrac{8 m_{B_c}}{3 m_{J/\psi} t^3} (\Delta^{B_c \to J/\psi})^2  (m_N^2 - t)^2 \lambda_t^{1/2} \nonumber \\ 
&& \times [\lambda_t (2 m_N^2 + t) + 6 m_{J/\psi}^2 t (m_N^2 + 2t)] ,
\eea

\noindent being a function of the squared transfer momentum $t$.


\end{document}